\documentclass{jfm}
\usepackage{amsmath}
\usepackage{amsfonts}
\usepackage{amssymb}
\usepackage{color}

\definecolor{grey0}{gray}{0.8}
\definecolor{grey1}{gray}{0.75}
\definecolor{grey2}{gray}{0.5}
\definecolor{grey3}{gray}{0.35}
\definecolor{grey4}{gray}{0.35}
\definecolor{blue1}{rgb}{0.88,0.88,1}
\definecolor{blue2}{rgb}{0.83,0.83,0.95}
\definecolor{blue3}{rgb}{0.8,0.8,0.9}
\definecolor{blue4}{rgb}{0.73,0.73,0.85}
\definecolor{blue5}{rgb}{0.6,0.6,0.8}
\definecolor{blue6}{rgb}{0.35,0.35,0.6}
\definecolor{blue7}{rgb}{0.2,0.2,0.45}
\definecolor{blue8}{rgb}{0.2,0.2,0.3}

\definecolor{blue9}{rgb}{0.4,0.6,1.}
\definecolor{blue10}{rgb}{0.1,0.1,0.9}
\definecolor{blue11}{rgb}{0.,0.,0.5}
\definecolor{red1}{rgb}{1,0.6,0.4}
\definecolor{red2}{rgb}{0.9,0.,0.}
\definecolor{red3}{rgb}{0.5,0.,0.}
\definecolor{green1}{rgb}{0.7,0.95,0.7}
\definecolor{green2}{rgb}{0.3,0.8,0.3}
\definecolor{green3}{rgb}{0.,0.5,0.}

\begin{document}

\newtheorem{lemma}{Lemma}
\newtheorem{corollary}{Corollary}

\shorttitle{Granular rheology in bedload transport} 
\shortauthor{R. Maurin et al} 

\title{Dense granular flow rheology in turbulent bedload transport}

\author
 {
Raphael Maurin\aff{1}
  \corresp{Present adress: Universit\'e de Toulouse, INPT, UPS, IMFT (Institut de M\'ecanique des Fluides de Toulouse), All\'ee Camille Soula, F-31400 Toulouse, France},
  Julien Chauchat\aff{2}
  \corresp{\email{julien.chauchat@grenoble-inp.fr}},
  \and 
  Philippe Frey\aff{1}
  }

\affiliation
{
\aff{1}
Univ. Grenoble Alpes, Irstea, UR ETGR, 2 rue de la Papeterie-BP 76, F-38402 St-Martin-d'H\`eres, France
\aff{2}
Univ. Grenoble Alpes, LEGI, G-INP,CNRS, F-38000 Grenoble, France
}

\maketitle

\begin{abstract}
The local granular rheology is investigated numerically in turbulent bedload transport. Considering spherical particles, steady uniform configurations are simulated using a coupled fluid-discrete-element model. The stress tensor is computed as a function of the depth for a series of simulations varying the Shields number, the specific density and the particle diameter. The results are analyzed in the framework of the $\mu(I)$ rheology and exhibit a collapse of both the shear to normal stress ratio and the solid volume fraction over a wide range of inertial numbers. Contrary to expectations, the effect of the interstitial fluid on the granular rheology is shown to be negligible, supporting recent work suggesting the absence of a clear transition between the free-fall and turbulent regime. In addition, data collapse is observed up to unexpectedly high inertial numbers $I\sim2$, challenging the existing conceptions and parametrization of the $\mu(I)$ rheology. Focusing upon bedload transport modelling, the results are pragmatically analyzed in the $\mu(I)$ framework in order to propose a granular rheology for bedload transport. The proposed rheology is tested using a 1D volume-averaged two-phase continuous model, and is shown to accurately reproduce the dense granular flow profiles and the sediment transport rate over a wide range of Shields numbers. The present contribution represents a step in the upscaling process from particle-scale simulations toward large-scale applications involving complex flow geometry. 
\end{abstract}

{\noindent \textbf{Key words:} Sediment transport, granular media, rheology}

\section{Introduction}

Among the different regimes of sediment transport, bedload transport is of major importance as it represents an important contribution to river morphology evolution. Accordingly, prediction of the sediment transport rate under turbulent bedload conditions is fundamental for preventing environmental risks associated with floods and scouring. From a physical point of view, bedload transport corresponds to the dynamic response of a granular bed submitted to a fluid shear stress. In contrast to suspended load, bedload is defined as the part of the sediment load occuring close to the granular bed in which particles are in permanent or intermittent contact with the bed \citep{Fredsoe1992}. 

The sediment transport rate is usually quantified in dimensionless form using the so-called Einstein number \citep{Einstein1942}, $Q_s^* = Q_s/[d\sqrt{(\rho^p/\rho^f -1)gd}]$, where $Q_s$ represents the volumetric sediment transport rate per unit width, $\rho^p$ and  $\rho^f$ are the particle and fluid densities,  $g$ is the acceleration of gravity and $d$ is the particle diameter. In the hydraulic community the classical way to model bedload transport consists of assuming a power law relation between the Einstein number and the Shields number $\theta$. The latter corresponds to the ratio of the traction force exerted by the fluid at the bed $\tau^f\ d^2$ and the buoyant weight of a single particle $(\rho^p -\rho^f)gd^3$, i.e. $\theta = \tau^f/[(\rho^p -\rho^f)gd]$. Given the lack of accuracy of the classical formulae (e.g. \citet{MPM1948}), renewed attention has recently been given to the granular phase behaviour in bedload transport \citep{Frey2009,Frey2011} from both an experimental \citep{Mouilleron2009,Hergault2010,Lajeunesse2010,Frey2014,Aussillous2013,Houssais2015} and a numerical point of view \citep{Duran2012,Ji2013,Kidanemariam2014,Maurin2015}. \\

The granular rheology characterises the response of the granular medium, in terms of deformation rate, to a given external stress and vice-versa. It is usually given as the relation between the granular stress and strain rate tensors. In bedload transport, the granular rheology governs both the response to, and the interaction with the fluid shear stress. Its knowledge is required for two-phase flow models in which both the fluid and solid phases are considered as continuous (Eulerian-Eulerian) (e.g. \citet{Aussillous2013}). As such it represents an important step in the upscaling of sediment transport processes from Eulerian-Lagrangian models to Eulerian-Eulerian models, which enable numerical simulations of larger-scale problems and/or complex flow forcings.  \\

Under bedload transport conditions, the granular medium experiences all of the granular flow regimes, from quasi-static in the sediment bed to dilute rapid granular flows in the upper sediment transport layer. The dilute rapid and dense granular flow regimes have mainly been described using the kinetic theory of granular flows and the $\mu(I)$ rheology respectively. The former is based on the analogy of dilute granular flows with molecular gases, assuming binary collisions. It has been proven to accurately describe rapid dilute granular flows in different configurations \citep{Campbell1990,Goldhirsch2003}, and has recently been extended to the dense granular flow regime \citep{Jenkins2006,Jenkins2007,Berzi2011}. Alternatively, the $\mu(I)$ local rheology accurately describes the dense granular flow regimes and is based on the dimensional analysis of the simple shear configuration \citep{GDRMidi2004,Forterre2008,Jop2015}. In the latter case, the unique dimensionless number controlling the system is the so-called inertial number \citep{DaCruz2005}: 
\begin{equation}
\label{Ieq}
I = \dot{\gamma} d \sqrt{\rho^p/P^p},
\end{equation} 
with $P^p$ the confining granular pressure and $\dot{\gamma}$ the granular shear rate. This dimensionless number can be interpreted as the ratio between a macroscopic time scale of deformation $t_{macro} = 1/\dot{\gamma}$ and a microscopic time scale of rearrangement $t_{micro} = d /\sqrt{P^p/\rho^p}$. From the dimensional analysis, the dimensionless solid volume fraction $\phi$ and shear to normal granular stress ratio $\mu = \tau^p/P^p$ are unique functions of the inertial number $I$. From experiments and numerical simulations, they can be expressed as \citep{DaCruz2005,Jop2006}
\begin{equation}
\frac{\tau^p}{P^p} = \mu(I) = \mu_s + \frac{\mu_2-\mu_s}{1+I_0/I},
\label{muIeq}
\end{equation}
\begin{equation}
\phi(I) = \phi^{max} - a I,
\label{phiIeq}
\end{equation}
where $\phi^{max}$ is the maximum packing fraction, $\mu_s$ is the static effective granular friction coefficient, and $\mu_2$, $I_0$ and $a$ are phenomenological constants. By combining the different expressions (eq. \ref{Ieq}-\ref{phiIeq}), the granular stress tensor can be expressed as a function of the shear rate, defining the granular rheology.\\

Based on the simple shear configuration, the $\mu(I)$ rheology has been extended to account for the presence of interstitial fluids \citep{CourrechDuPont2003,Cassar2005}. Considering rearrangement time scales according to the dominant mechanism, three regimes can be defined by introducing two dimensionless numbers \citep{CourrechDuPont2003,Cassar2005,Andreotti2013}, with $C_D$ the drag coefficient,
\begin{equation}
St = \frac{d\sqrt{\rho^p P^p}}{\eta^f},
\label{Steq}
\end{equation}
\begin{equation}
r = \sqrt{\frac{\rho^p}{\rho^f C_D}},
\label{req}
\end{equation}
the free-fall regime ($St>>1$, $r>>1$) corresponding to negligible influence of the interstitial fluid (i.e. dry granular media),  the viscous regime ($St<<1$, $r<<1$) corresponding to a rearrangement time scale dominated by viscous drag and the turbulent regime ($St>>1$, $r<<1$) corresponding to a rearrangement time scale dominated by fluid inertial effects. In the different regimes, the results therefore scale with the dry inertial number $I_{dry} =\dot{\gamma} d\sqrt{\rho^p/P^p}$, the turbulent one $I_{turb} =\dot{\gamma} d\sqrt{\rho^f C_D/P^p}$ and the viscous one $I_{visc} = \eta^f \dot{\gamma}/P^p$ respectively. This approach has been applied with some success to various complex immersed configurations from  avalanches \citep{CourrechDuPont2003,Cassar2005,Doppler2007}, to granular collapses \citep{Rondon2011,Izard2014}, annular shear cells \citep{Boyer2011,Trulsson2012} and sediment transport \citep{Ouriemi2009,RevilBaudard2013,Aussillous2013,Chiodi2014}. \\

To the best of our knowledge, there are only a few contributions in the literature on the dense granular flow rheology in the context of bedload transport. \citet{Ouriemi2009} and \citet{Aussillous2013} have studied laminar bedload in closed conducts using refractive index-matching experiments and theoretical two-phase continuous models. Assuming a constant solid volume fraction and a limited transitional layer from dense to dilute granular flow, the authors have shown that the $\mu(I)$ rheology gives excellent agreement with experimental data. It is worth noting that in the laminar case, the lag between the fluid and solid velocities is negligible \citep{Mouilleron2009,Aussillous2013}, the two phases being tightly coupled. In intense turbulent bedload transport, termed sheet flow, the transitional layer from dense to dilute granular flows and the fluid-particle velocity lag are more important \citep{Sumer1996,Cowen2010,RevilBaudard2015}. This regime has been studied experimentally \citep{Capart2011} and numerically using the two-phase flow continuous modelling framework with the kinetic theory of granular flows \citep{Jenkins1998,Hsu2004} and the $\mu(I)$ rheology \citep{RevilBaudard2013,Chauchat2015}. While the agreement of the numerical simulations with the available experimental data gives some credit to both granular rheologies, recent experimental investigations by \citet{RevilBaudard2015} showed that the solid volume fraction profiles observed were substantially different from the predicted ones. The observed effective friction coefficient also differs importantly from the classical values used in the $\mu(I)$ rheology, possibly due to the strong intermittency induced by the turbulent coherent structures \citep{RevilBaudard2015}. At the transition from sheet flows to debris flow, the granular rheology has been used to model steep slope configurations \citep{Armanini2005,Larcher2007,Berzi2008,Armanini2014}. Combining experimental particle tracking at the wall \citep{Armanini2005,Larcher2007} and numerical analysis \citep{Berzi2008,Armanini2014}, both the $\mu(I)$ rheology \citep{Berzi2008} and hybrid $\mu(I)$--kinetic theory models \citep{Armanini2014} have been shown to accurately describe the granular behaviour in this configuration.\\

The literature review underlines the relevance of both the kinetic theory and the dense granular flow rheology for bedload transport modelling. However, the numerical studies are restricted to continuous two-phase analysis and the experimental evaluations of the granular stress tensor rely on solid volume fraction measurements that are highly uncertain. In addition, while direct numerical simulations (DNS) have been performed at the particle scale in the laminar regime \citep{Kidanemariam2014}, this type of numerical model is not affordable for turbulent bedload transport due to the typical values of bulk Reynolds numbers explored. Therefore, to go beyond the existing two-phase continuous numerical works, the present paper analyses the local granular rheology in turbulent bedload transport using a coupled fluid-discrete-element model (DEM) with a volume-averaged fluid description \citep{Maurin2015}. This approach permits local computation of the granular stress tensor and the granular shear rate, bringing new insights into the local granular behaviour in turbulent bedload transport. Based on these Eulerian-Lagrangian simulations a dense granular flow rheology for bedload transport is proposed and further tested in a two-phase continuous (Eulerian-Eulerian) model. Besides, bedload transport configurations enable local analysis of the granular rheology in complex immersed granular flows, and the DEM results are shown to be of interest for granular media in general. \\

The paper is structured as follows. First, the coupled fluid-DEM and two-phase continuous models are briefly presented (section \ref{modelFormulation}). Then, fluid-DEM simulations are analyzed in the framework of the dense granular flow $\mu(I)$ rheology (section \ref{Results}). Finally, extending the simulations to realistic conditions for bedload transport, a parametrization of the $\mu(I)$ rheology is proposed from the DEM results, and is tested with the two-phase continuous model (section \ref{graRheoBed}).

\section{Model formulation}
\label{modelFormulation}

The Eulerian-Lagrangian and Eulerian-Eulerian models are based on the same volume-averaged two-phase flow equations for the fluid phase. The solid phase is modelled on the one hand as a continuum (Eulerian-Eulerian) and on the other hand using a 3D DEM in which each particle motion is computed explicitly (Eulerian-Lagrangian). Both two-phase flow models have already been described and validated with experiments \citep{Maurin2015,Chauchat2015} and only a brief description will be given. The reader is referred to the abovementioned references for a complete description of the model formulations and validations. \\ 

Considering steady uniform conditions, the problem is unidirectional so that the average fluid and solid velocities depend only on the wall-normal direction and reduce to their streamwise components $\left<\mathbf{u}\right>^f = \left<u_x\right>^f(z) \ \mathbf{e_x}$ and  $\left<\mathbf{v}^p\right>^s = \left<v_x^p\right>^s(z) \ \mathbf{e_x}$.  Therefore, the streamwise and wall-normal volume-averaged fluid phase momentum balances are given by \citep{Anderson1967,Jackson2000,RevilBaudard2013}
\begin{equation}
0 = \frac{\partial S_{xz}^f}{\partial z} + \frac{\partial R_{xz}^f}{\partial z} + \rho^f (1-\phi) g \sin \alpha - n \left<f_x\right>^p,
\label{fluidXZ}
\end{equation}
\begin{equation}
0 = - \frac{\partial P^f}{\partial z} + \rho^f (1-\phi)  g \cos \alpha -  n \left<f_z\right>^p,
\label{fluidZZ}
\end{equation}
where $\sigma^f_{ij}= -P^f \delta_{ij} + S^f_{ij}$ is the volume-averaged effective viscous stress tensor, $R^f_{xz}$ is the Reynolds shear stress tensor, $\phi$ is the solid phase volume fraction, $\left<f_k\right>^p$ is the volume-averaged fluid-particle interaction force, $n = \phi/(\pi d^3/6)$ is the number density of particles and $\alpha$ is the channel inclination angle. Omitting the model-dependent fluid-particle interaction force, the solution of the fluid momentum balance requires closure laws for the viscous shear stress tensor and the Reynolds stress tensor. \\

Considering a Newtonian fluid, the viscous shear stress is classically expressed as
\begin{equation}
S_{xz}^f = \rho^f (1-\phi) \nu^f \frac{d \left<u_x\right>^f}{dz},
\label{viscousStressTensorClosure}
\end{equation}
with $\nu^f$ the clear fluid kinematic viscosity and $\left<u_x\right>^f$ the volume-averaged streamwise fluid phase velocity. 
The Reynolds shear stress is based on the eddy viscosity concept ($\nu^t$) using a mixing length formulation: 
\begin{equation}
 R_{xz}^f =  \rho^f ~ \nu^t \frac{d \left<u_x\right>^f}{dz} \ \ \text{with}\ \ \nu^t = (1-\phi)  \ l_m^2 \left|\frac{d \left< u_x \right>^f}{dz}\right|,
\label{ReynoldsStressTensorClosure}
\end{equation}
where the mixing length is taken similarly to \citet{Li1995} as
\begin{equation}
l_m(z) = \kappa \int_0^z{\frac{\phi^{max} - \phi(\zeta)}{\phi^{max}} ~d\zeta},
\label{mixingLength}
\end{equation}
with $\kappa=0.41$ the von Karman constant. The formulation adopted allows one to recover the law of the wall \citep{Prandtl1926} in clear fluid, while the turbulence is completely damped inside the granular bed at maximum packing fraction ($\phi^{max}$).

\subsection{Eulerian-Lagrangian model}

The Eulerian-Lagrangian model is based on the explicit solution of the dynamic equation for each individual particle using the 3D DEM open-source code YADE \citep{YADEDEM2015}. The DEM solution is spatially averaged and explicitly coupled with the fluid phase momentum balance equations (\ref{fluidXZ}-\ref{fluidZZ}) through the drag term $n<f^D_x>^p$ and the solid volume fraction $\phi$.

For each particle $p$ at position $\mathbf{x}^p$, the Newton equations are solved considering nearest-neighbours interactions: 
\begin{equation}
m \frac{d^2 \mathbf{x}^p}{d t^2} =  \sum_{k \in {\cal N}}\mathbf{f}_c^{pk}  + \mathbf{f}_{ext} =  \sum_{k \in {\cal N}}\mathbf{f}_c^{pk} + \mathbf{f}_g^p + \mathbf{f}^p_b +  \mathbf{f}^p_D
\label{DEM}
\end{equation}
where the sum of the contact forces $\mathbf{f}_{c}^{pk}$ is made over the ensemble of nearest neighbours ${\cal N}$, $\mathbf{f}_g^p$ is the gravity force, $\mathbf{f}_b^p$ is the buoyancy force and $\mathbf{f}_D^p$ is the 3D drag force applied by the fluid on particle $p$. For each contact, the contact force is computed explicitly using the classical spring-dashpot contact law \citep{Schwager2007}:
\begin{subequations}
 \begin{align}
&F_n = -k_n \delta_n - c_n \dot{\delta_n}\\[5pt]
&F_t = -{\rm min}(k_s \delta_t,\mu^p F_n),
 \end{align}
\label{fluctEq}
\end{subequations}
where $F_n$ and $F_t$ are the normal and tangential contact forces between particle $p$ and $k$, $\delta_n$ and $\delta_t$ are the normal and tangential overlaps, $k_n$ and $k_s$ are the normal and tangential contact stiffnesses, $c_n$ is the normal viscous damping and $\mu^p$ is the tangential friction coefficient. This contact law is well suited for granular flow analysis and allows one to define a unique restitution coefficient $e_n$ \citep{Schwager2007}. Similarly, the rotation of the particles is solved from the Newton equations of motion. \\

The interaction with the fluid phase is restricted to buoyancy and drag forces: 
\begin{equation} 
\mathbf{f}_{b}^p  =  -\frac{\pi d^3}{6}\mathbf{\nabla}P^f,
\label{generalisedBuoyancy}
\end{equation}
\begin{equation}
\displaystyle \mathbf{f}_{D}^p = \frac{1}{2}\rho^f \frac{\pi d^2}{4} ~ C_D ~ \left|\left| \left<\mathbf{u}\right>^f_{\mathbf{x^p}} - \mathbf{v^p} \right|\right|\left(\left<\mathbf{u}\right>^f_{\mathbf{x^p}} - \mathbf{v^p}\right),
\label{drag}
\end{equation}
where the average fluid velocity and the fluid pressure are taken at the center of particle $p$. The drag coefficient $C_D$ depends on the particle Reynolds number $\Rey_p = \left|\left| \left<\mathbf{u}\right>^f_{\mathbf{x^p}} - \mathbf{v^p} \right|\right| d/\nu^f$ and takes into account hindrance effects \citep{Dallavalle1948,Richardson1954}: $C_D = \left(0.4+ 24.4/Re_p\right) (1-\phi)^{-3.1}$.\\

Considering steady uniform turbulent flows, the buoyancy force reduces to its wall-normal component and equation (\ref{fluidZZ}) leads to a hydrostatic fluid pressure distribution, the wall-normal average drag force being negligible. The solution of the streamwise fluid phase momentum balance (eq. \ref{fluidXZ}) requires the evaluation of both the spatially averaged solid phase volume fraction and the momentum transmitted from the fluid to the particles through drag forces: 
\begin{equation}
n<f^D_x>^s = \frac{\phi}{\pi d^3/6}  <f^D_x>^s = \frac{3}{4}~\frac{\phi ~ \rho^f}{d} \left<  C_D ~ \left|\left| \left<\mathbf{u}\right>^f - \mathbf{v^p} \right|\right| ~ \left( \left<u_x\right>^f - v^p_x\right) \right>^s.
\label{averageDrag1final}
\end{equation}
In the DEM processing, the solid phase averaging $<\bullet>^s$ is used instead of the particle phase averaging $<\bullet>^p$. While the two are identical provided that there is scale separation \citep{Anderson1967,Jackson2000}, the importance of the wall-normal gradients in bedload transport requires a weighting function length scale lower than the particle diameter in order to define an independent averaging \citep{Maurin2015,MaurinPhD}. Therefore, a small wall-normal weighting function length scale has been adopted (typically $d/30$), and this choice has been validated through an experimental comparison \citep{Maurin2015}.\\

The 3D DEM and the fluid model are solved as transient problems applying a fixed bottom boundary condition for both the fluid ($\left<\mathbf{u}\right>^f(z = 0) = 0$) and the particle phase (fixed random particles) and imposing the position of the water free-surface ($d \left<u_x\right>^f/dz(z = h) = 0$). In order to achieve a stable integration, the DEM time step is bounded by the propagation time of the fastest wave over a particle diameter \citep{MaurinPhD,Maurin2015}. The fluid resolution time step corresponds to a typical characteristic evolution time scale of the granular medium and is taken much larger than the DEM one \citep{Maurin2015}: $\Delta t_f = 10^{-2}s$ with respect to $\Delta t_p \sim O(10^{-4}-10^{-5}) \ s$. Therefore, the coupling between the fluid and granular phases ensures momentum conservation of the system on average. The model has been compared with experiments and has shown its ability to describe accurately the granular depth structure in turbulent bedload transport \citep{Maurin2015}.

\subsection{Eulerian-Eulerian model}

The Eulerian-Eulerian model is based on the numerical solution of the fluid phase momentum balance equations (eqs \ref{fluidXZ} and \ref{fluidZZ}) coupled with the spatially averaged granular phase momentum balance equations in the streamwise and wall-normal directions. For steady uniform flow conditions the granular phase momentum equations read \citep{Chauchat2015}
\begin{equation}
0 = \frac{\partial \tau_{xz}^p}{\partial z} + \rho^p \phi  g \sin \alpha + n \left<f_x\right>^p,
\label{partXZ}
\end{equation}
\begin{equation}
0 = - \frac{\partial P^p}{\partial z} + \rho^p \phi  g \cos \alpha +  n \left<f_z\right>^p,
\label{partZZ}
\end{equation}
where $\tau_{xz}^p$ is the spatially-averaged granular phase shear stress. Like in the Eulerian-Lagrangian model, the buoyancy is applied along the wall-normal direction and leads to hydrostatic pressure distribution for the granular phase. In the continuous formalism the spatially averaged streamwise drag force is given by
\begin{equation}
n\left<\mathbf{f}^D\right>^s = \frac{3}{4} \frac{\phi \rho^f}{d} ~ C_D \left|\left| \left<\mathbf{u}\right>^f - \left<\mathbf{v^p}\right>^s + \mathbf{u_{d}} \right|\right|\left(\left<\mathbf{u}\right>^f -  \left<\mathbf{v^p}\right>^s + \mathbf{u_{d}}\right),
\label{drag}
\end{equation}
where the extra term $\mathbf{u_{d}}$ is the drift velocity which represents the dispersion effect due to the averaged fluid-particle velocity fluctuations. This term is responsible for sediment transport suspension and allows one to reproduce the Rouse profile provided that the wall normal component is taken as: $(u_{d})_z= -  \nu^t/\phi \ {\rm d} \phi / {\rm d} z$ \citep{Chauchat2015}. The streamwise component is taken as $(u_d)_x=0$ in this paper.

The granular shear stress tensor is modeled using the $\mu(I)$ rheology presented in the introduction. Combining equations (\ref{Ieq}) and (\ref{muIeq}), the shear and normal components of the granular stress tensor can be expressed as
\begin{equation}
\displaystyle P^p = \frac{1}{I^2} \rho_p \ d^2 \ \left(\frac{d \left<v^p_x\right>^s}{d z}\right)^2,
\end{equation}
\begin{equation}
\displaystyle \tau^p = \frac{\mu(I)}{I^2} \rho_p \ d^2 \ \left(\frac{d \left<v^p_x\right>^s}{d z}\right)^2.
\end{equation}
Inverting equation \ref{phiIeq}, the inertial number is obtained as a function of the solid volume fraction and the two equations can be directly expressed as a function of the solid volume fraction and the average solid velocity gradient. \\

The numerical algorithm is based on a transient solution of the governing equations (\ref{fluidXZ}; \ref{fluidZZ}; \ref{partXZ} and \ref{partZZ}), together with the continuity equation for the solid phase. Fixed boundary conditions are imposed at the bottom of the sample at $z=0$ ($\left<\mathbf{u}\right>^f = 0$) where the granular phase is considered to be at rest ($\left<\mathbf{v}^p\right>^s = 0$, $d\phi/dz = 0$), while the positions of the water and granular free-surfaces are imposed in $z=h$ ($d\left<u_x\right>^f/dz = 0$, $\left<u_z\right>^f =  0$, $\left<v_z^p\right>^s=0$,  $d\phi/dz= 0$). 
The governing equations are discretised using a finite volume technique for the mass conservation equation and a finite difference technique for the momentum balance. A staggered grid is used with the velocities located at the cell face and the scalar quantities (e.g. volume fractions or viscosities) located at the cell center. For the pressure-velocity coupling a projection method is used. The numerical schemes are Euler implicit for the time derivative, upwind for the advection terms and central difference for the diffusion terms. For additional informations on the numerical model, the reader is referred to \citet{Chauchat2013} and \citet{Chauchat2015}.

\section{Results and methodology}
\label{Results}
\begin{figure}
  \centerline{  \includegraphics[width=\textwidth]{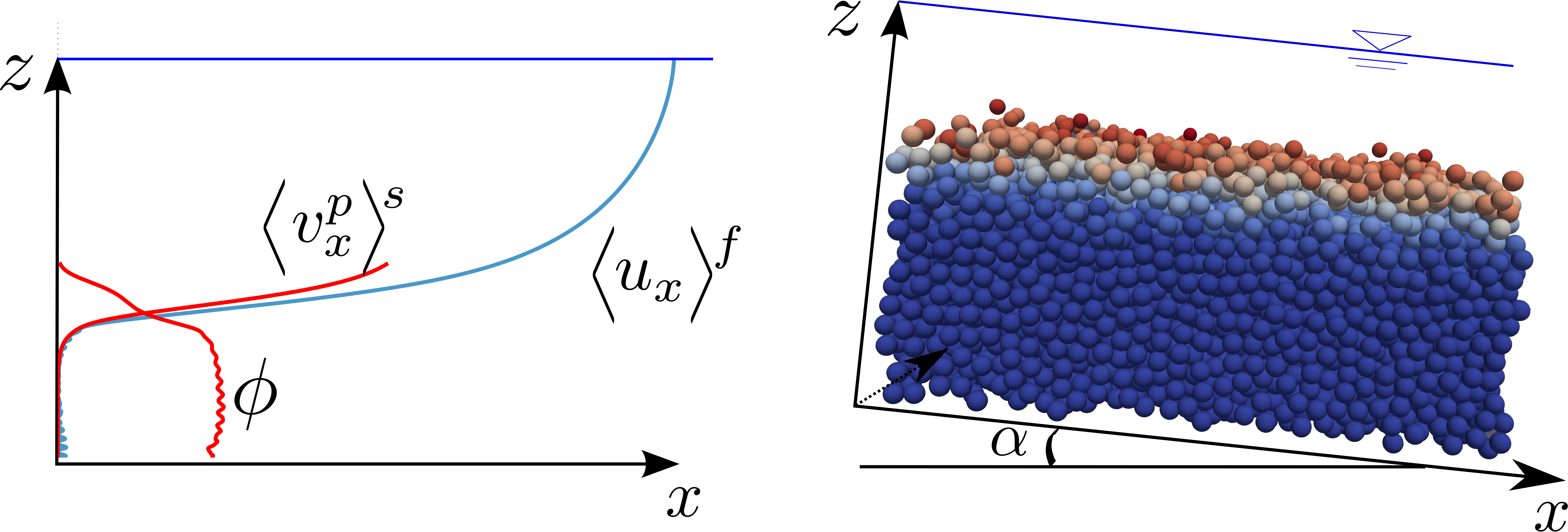} }
\caption{\label{figSituation} Scheme of the numerical setup and its equivalent average unidirectional picture with typical fluid velocity $\left<\mathbf{u}^f\right> = \left<u_x^f\right>(z) \ \mathbf{e_x}$, solid volume fraction $\phi$, and solid velocity $\left<\mathbf{v}\right>^s = \left<v_x\right>^s(z) \ \mathbf{e_x}$ depth profiles. The inclined 3D bi-periodic granular description is coupled with a unidirectional fluid momentum balance using imposed fixed random bottom boundary condition and water free-surface position. The particle color is representative of the velocity intensity. }
\end{figure}

The numerical setup of the Eulerian-Lagrangian model is presented in figure \ref{figSituation}. It consists of a channel flow tilted with an inclination angle $\alpha = 0.05$ rad, partially filled with monodisperse spherical particles of diameter $d$. The three-dimensional granular sample is biperiodic (streamwise and spanwise) and the periodic cell size is taken as $l_x = l_y = 30d$ to ensure statistical convergence of the spatial averaging operator \citep{Maurin2015}. A 10-15 diameter thick granular layer is deposited under gravity over a rough random bottom, and the system evolves under gravity with a fixed water free-surface elevation. The water flow is turbulent ($\Rey = U^f h/\nu^f \sim 10^4$), hydraulically rough ($\Rey_p = U^f d/\nu^f \sim 10^3$) and supercritical ($Fr = U^f/\sqrt{g h} \gtrsim 1$), with $U^f$ the average fluid velocity within the water depth $h$. The results are independent of the granular bottom boundary conditions in the range of parameters investigated \citep{MaurinPhD} and the simulations are performed in the rigid grain limit \citep{Roux2002} with a tangential stiffness set to half the normal one. The friction coefficient is set to a realistic value for glass beads, $\mu^p = 0.4$, and the restitution coefficient is taken as $e_n = 0.5$ to account for the lubrication effect consistently with the experimental validation \citep{Maurin2015}.\\

\begin{table}
 \begin{center}
  \begin{tabular}{cccccccc}
$\theta^*$ & $\Rey$ & $\Rey_p$ & $\rho^p/\rho^f -1$ & $Fr$ &  $S^*$ & $St$ & $r$\\
$[0.04,0.6]$ & $10^3-10^6$ & $10^3-10^4$ & $[0.375,1.5]$ & $\gtrsim 1$  & $[2~, 5]$ & $10^2-10^4$ & $0.1-2$
  \end{tabular}
  \caption{Main characteristic dimensionless numbers of the configurations considered.}\label{dimNumb}
 \end{center}
\end{table}

In order to define properly an equivalent continuous medium for the granular phase, the analysis is restricted, in a first approach, to cases with a non-negligible number of particle layers in motion. The range of Shields numbers investigated has been chosen between $\theta^* \sim 0.2$ and $\theta^* \sim 0.6$, to prevent suspended load and stay in the bedload regime by keeping high suspension numbers ($S^* = w_s/u_* \in [2~, 5]$). For mountain stream bedload, this represents intense rare events with high transport capacity. It can also be seen as a limit case of sheet flow without suspension. In order to investigate the scaling laws, the Shields number, the specific density and the particle diameter are varied. For each parameter, three values are considered starting from a realistic case for intense bedload transport ($\theta^* \sim 0.4$, $\rho^p/\rho^f -1 = 1.5$, $d = 6$mm), corresponding to the particle density and diameter used for the experimental validation \citep{Maurin2015}. The specific density is lowered starting from $\rho^p/\rho^f-1 = 1.5$, considering that experimental data are often made with plastic in sheet flows (e.g. \citet{Capart2011,RevilBaudard2015}). The main characteristic dimensionless numbers of the problem are shown in table \ref{dimNumb} and the exact parameters investigated are given in table \ref{tableRheoParam}. For each run, once the system is at steady state, the data measured every $0.1s$ are averaged over time for $300s$ for post-processing.

\begin{table}
 \begin{center}
  \begin{tabular}{lccccc}
Run & $\rho^p/\rho^f - 1 $ & $d (mm)$ & $\theta^*$ & symbol\\
r0d3s1  	& 0.375	& 3    & 0.214   &  $\color{green1}{\bullet}$\\
r0d3s2  	& 0.375	& 3    & 0.444   &  $\color{green2}{\bullet}$\\
r0d3s3  	& 0.375	& 3    & 0.611   &  $\color{green3}{\bullet}$\\
r0d6s1  	& 0.375	& 6    & 0.215   &  $\color{green1}{+}$\\
r0d6s2  	& 0.375	& 6    & 0.44	 &  $\color{green2}{+}$\\
r0d6s3  	& 0.375	& 6    & 0.598   &  $\color{green3}{+}$\\
r0d12s1		& 0.375 & 12   & 0.216	 &  \textcolor{green1}{x}\\  
r0d12s2		& 0.375 & 12   & 0.434	 &  \textcolor{green2}{x}\\
r0d12s3		& 0.375 & 12   & 0.593	 &  \textcolor{green3}{x}\\
r1d3s1  	& 0.75 	& 3    & 0.188   &  $\color{red1}{\bullet}$\\
r1d3s2  	& 0.75 	& 3    & 0.378   &  $\color{red2}{\bullet}$\\
r1d3s3  	& 0.75 	& 3    & 0.593   &  $\color{red3}{\bullet}$\\
r1d6s1  	& 0.75 	& 6    & 0.193   &  $\color{red1}{+}$\\
r1d6s2  	& 0.75 	& 6    & 0.381   &  $\color{red2}{+}$ \\
r1d6s3  	& 0.75 	& 6    & 0.598   &  $\color{red3}{+}$ \\
r1d12s1 	& 0.75 	& 12   & 0.191   &  \textcolor{red1}{x} \\
r1d12s2 	& 0.75 	& 12   & 0.379   &  \textcolor{red2}{x}\\
r1d12s3 	& 0.75 	& 12   & 0.596   &  \textcolor{red3}{x}\\
r2d3s1  	& 1.5 	& 3    & 0.205   &  $\color{blue9}{\bullet}$\\
r2d3s2  	& 1.5 	& 3    & 0.443   &  $\color{blue10}{\bullet}$\\
r2d3s3  	& 1.5 	& 3    & 0.694   &  $\color{blue11}{\bullet}$\\
r2d6s1  	& 1.5 	& 6    & 0.205   &  $\color{blue9}{+}$\\
r2d6s2  	& 1.5 	& 6    & 0.455   &  $\color{blue10}{+}$\\
r2d6s3  	& 1.5 	& 6    & 0.692   &  $\color{blue11}{+}$\\
r2d12s1 	& 1.5 	& 12   & 0.21	 &  \textcolor{blue9}{x}\\
r2d12s2 	& 1.5 	& 12   & 0.451   &  \textcolor{blue10}{x}\\
r2d12s3 	& 1.5 	& 12   & 0.696   &  \textcolor{blue11}{x}\\ 
  \end{tabular}
  \caption{Parameters of the simulations studied and symbol correspondence. The specific density, particle diameter, and Shields number have been varied. The latter is evaluated from the maximum of the turbulent shear stress, using $\tau^f = max(R_{xz}^f) = \rho^f u_*^2$.  Each specific density is associated with a color, which intensity reflects the Shields number. The symbol associated with the run is characteristic of the particle diameter.}\label{tableRheoParam}
 \end{center}
\end{table}

\subsection{Methodology}

In order to study the granular rheology, the granular stress tensor is computed from the 3D DEM results by applying the spatial averaging operator to the granular stress tensor. Therefore, $\left<\sigma_{ij}^p \right>^s $ is obtained for each slice of volume $V$ by computing \citep{Goldhirsch2010,Andreotti2013}
\begin{equation}
\left<\sigma_{ij}^p \right>^s = - P^p \delta_{ij} + \tau^p_{ij} = -\frac{1}{V} \sum_{\beta \in V} m^{\beta} v_i^{'\beta} v_j^{'\beta} - \frac{1}{V} \sum_{(m,n) \in V}f_{c,i}^{m,n} b_j^{m,n}, 
\label{stressTensor}
\end{equation}
where the sums are respectively over the particles and the contacts contained in the volume $V$, $v_k^{'\beta} = v_k^{\beta} - \left<v_k\right>$ is the $k^{th}$ component of the spatial velocity fluctuation associated with particle $\beta$ of mass $m^{\beta}$, $\mathbf{f}_c^{m,n}$ is the contact force applied by particle $m$ on particle $n$ and $\mathbf{b}^{m,n}$ is the branch vector from  particle $m$ to particle $n$.\\

\begin{figure}
  \centerline{  \includegraphics[width=\textwidth]{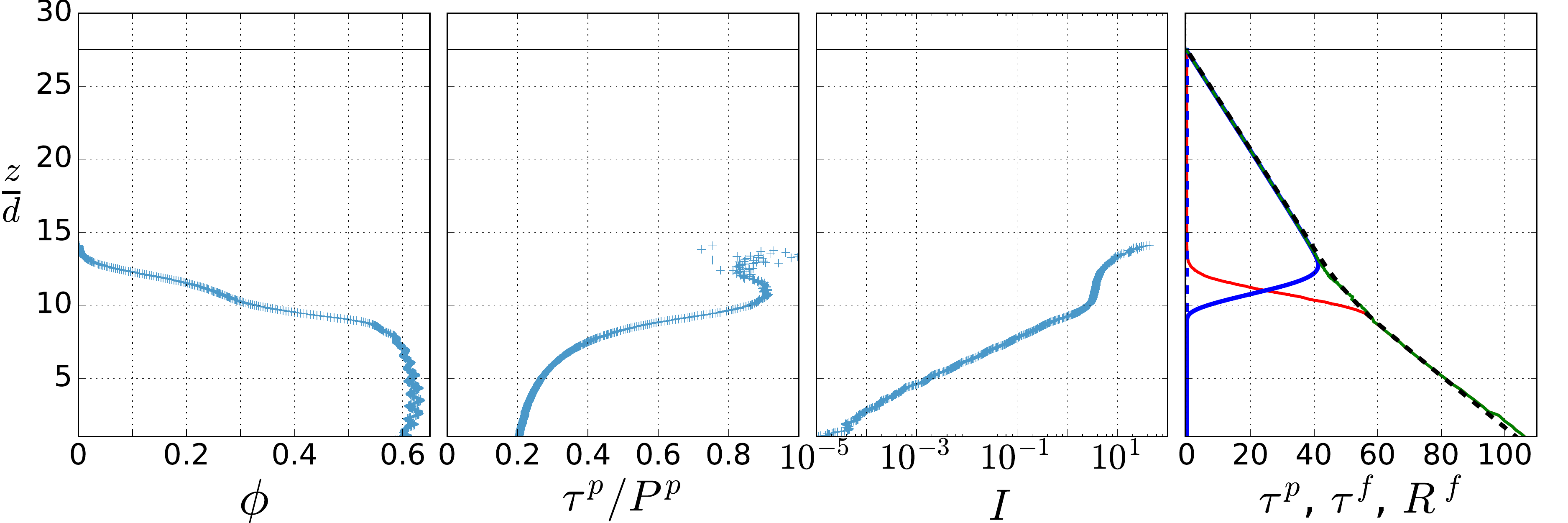} }
\caption{\label{methodo} Representative example (case r2d6s2 in table \ref{tableRheoParam}) of solid volume fraction, shear to normal granular stress ratio, inertial number, and total stress repartition depth profiles. The last panel shows the different components of the mixture momentum balance as a function of the depth, as derived in appendix \ref{appendixMomBal}: the granular shear stress (\textbf{\textcolor{red}{--}}), the Reynolds shear stress (\textbf{\textcolor{blue}{--}}), the viscous shear stress  (\textbf{\textcolor{blue}{- -}}), the sum of the three (\textbf{\textcolor{green}{--}}), and the slope contribution (last term of equation \ref{mixtureXZ_INT}, \textbf{\textcolor{black}{- -}}). The water free-surface elevation is represented in each different panel for reference (--).}
\end{figure}

Considering the unidirectional character of the problem, the weighting function of the spatial averaging operator extends over the whole width and length of the periodic cell. Therefore, one obtains for each run time-averaged depth profiles of the shear and normal granular stress components: $\tau^p(z) = \tau^p_{xz}(z)$ and  $P^p(z) = \tau^p_{zz}(z)$. Negligible stress asymmetry and compensated normal stress difference have been observed \citep{MaurinPhD}. Figure \ref{methodo} presents typical depth profiles for a given representative simulation (run r2d6s2 in table \ref{tableRheoParam}). The last panel shows the different components of the streamwise mixture momentum balance, as derived in appendix \ref{appendixMomBal}. Similarly to \citet{RevilBaudard2013} in the sheet flow regime, the viscous stress tensor contribution is found to be negligible in turbulent bedload transport. In addition, the momentum balance is closed at each elevation, as the streamwise projection of the gravity contribution is balanced by the sum of the other terms, showing the consistency of the model formulation and stress tensor evaluation. The depth profile of the inertial number (figure \ref{methodo}) shows that a very important range of inertial numbers (approximately five orders of magnitude) is sampled as a function of the depth in a single simulation. This is due to the nature of bedload transport in which the granular flow evolves from quasi-static in the bed to very dynamic at the granular free-surface. As the shear to normal stress ratio and the solid volume fraction vary accordingly throughout the depth (see figure \ref{methodo}), this allows us to obtain the granular rheology for a wide range of inertial numbers from a unique simulation. Performing simulations with variation of $\rho_p$, $d$ and $h$ the water depth, the results take the form of a set of $\mu(I)/\phi(I)$ curves sampling the parameter space. In the present paper, each simulation will be represented by a colored symbol, where the symbol is associated with the particle diameter $d$, its color is associated with the value of the specific density $\rho_p/\rho_f-1$ and its shading represents the Shields number (the darker the colour, the higher the Shields number) (see table \ref{tableRheoParam}).

\subsection{Results}

The influence of the interstitial fluid on the granular rheology varies between and within the different configurations investigated and one might expect transitions between the different regimes of the $\mu(I)$ rheology (free-fall, turbulent, viscous). While generally of minor importance, the choice for the computation of the Stokes and $r$ dimensionless numbers (eq. \ref{Steq} and \ref{req}) becomes crucial when approaching a transition between two regimes, as in the present analysis. In order to define accurately the different regimes and to impose that the transition from the viscous to the turbulent regime depends only on the particle Reynolds number ($Re_p = St/r$), the limit turbulent drag coefficient $C_D^{\infty} = 0.44$ should be considered and both turbulent and viscous drag forces should be taken without the contribution from hindrance effects. However, the classical picture of the $\mu(I)$ rheology associates the inertial number with a time-scale ratio representative of the local rearrangement process \citep{Andreotti2013}, which is affected by hindrance effects and the particle Reynolds number in the low-$Re_p$ limit of the turbulent regime ($Re_p \in [1,10^4]$). Focusing on the present case ($St>>1$ and $r\sim1$ so that $Re_p>>1$), the $r$ and turbulent inertial numbers have been computed from a constant drag coefficient taking into account the local hindrance effects $C_D(z) = C_D^{\infty} (1-\phi(z))^{-3.1}$, putting aside the considerations on the transition from the viscous to the turbulent regime. \\

\begin{figure}
  \centerline{\includegraphics[width=0.48\textwidth]{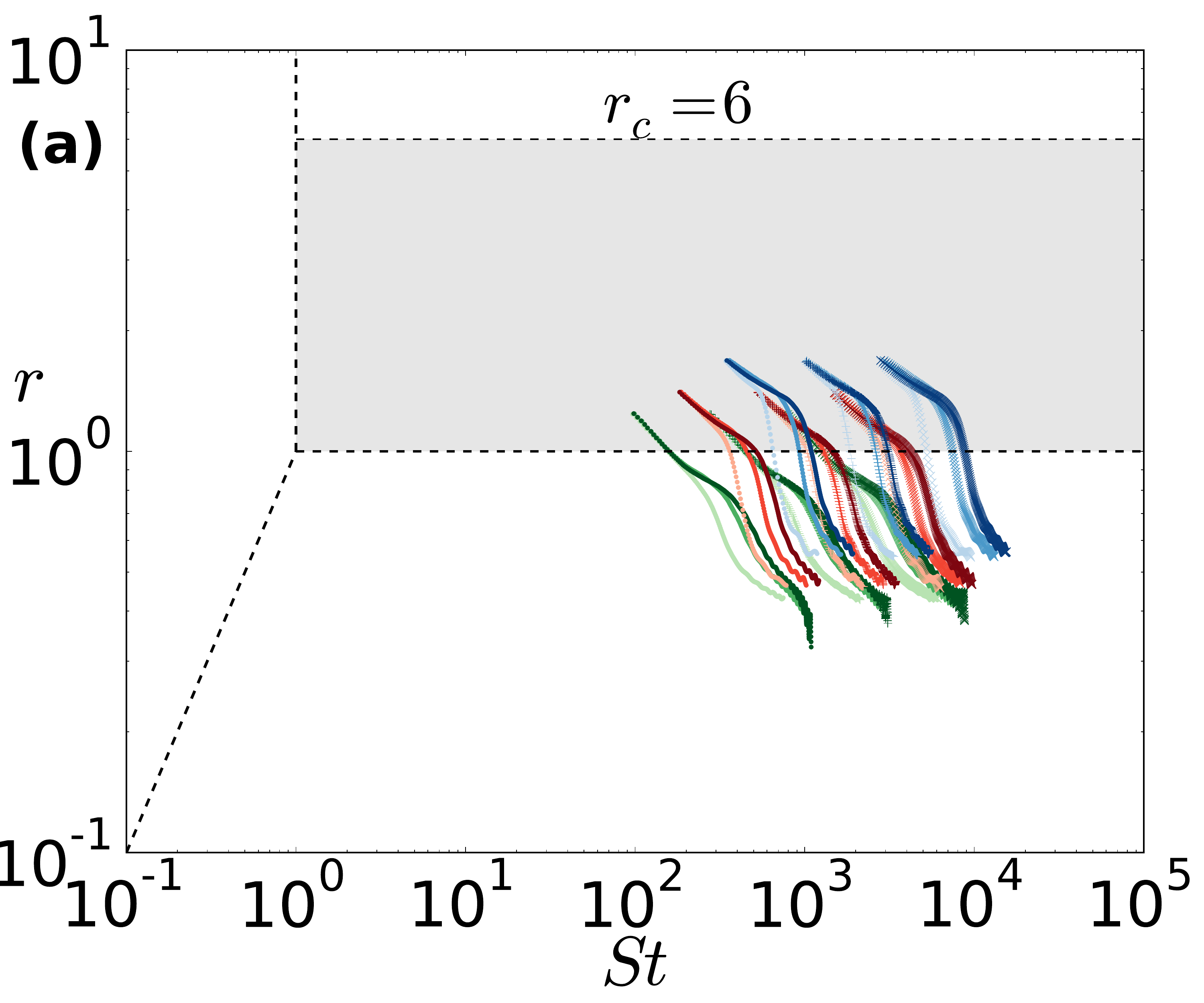}\includegraphics[width=0.48\textwidth]{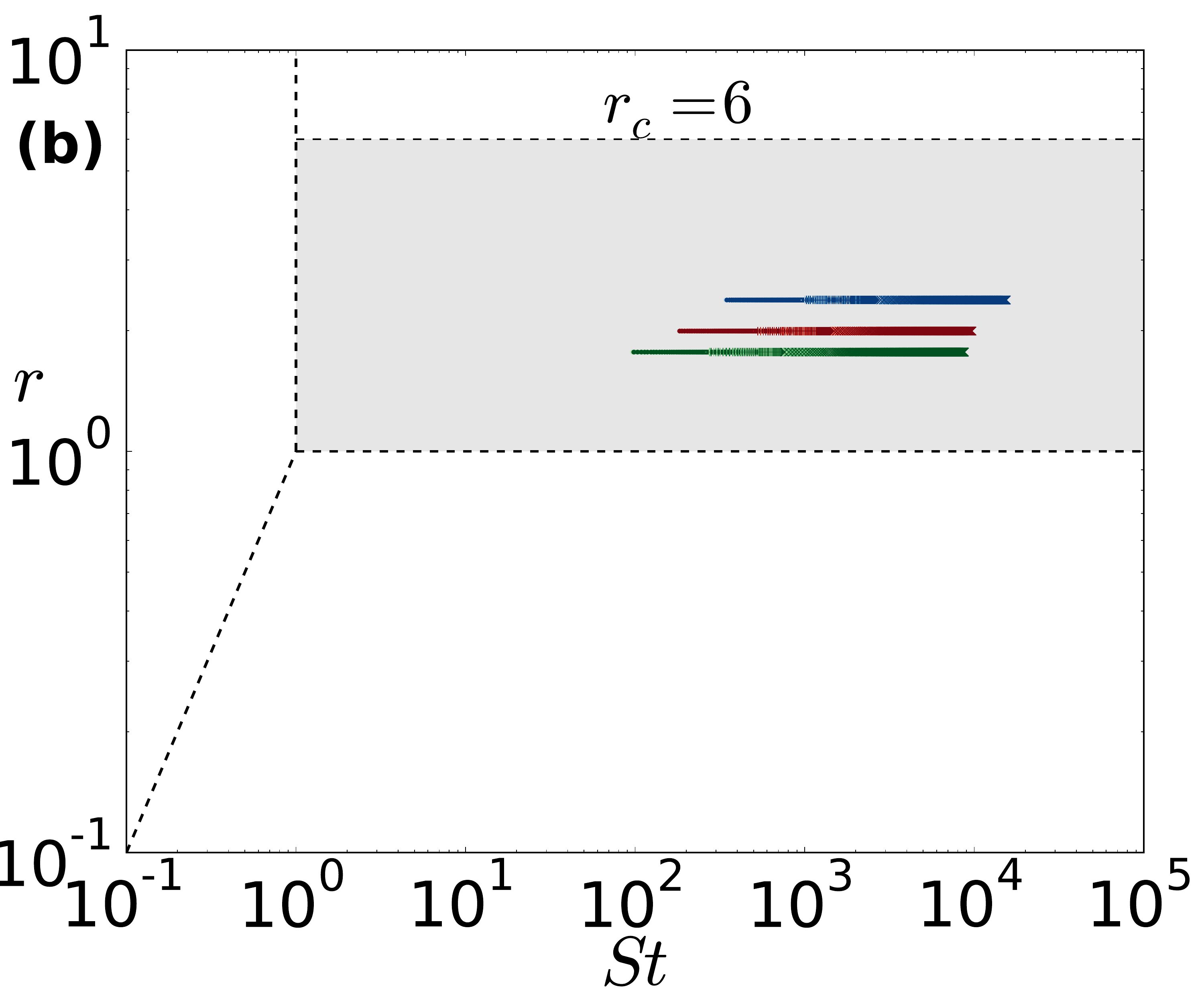}}
\caption{\label{rSt} Position of all the computed granular rheology points in the (St,r) plane, considering the drag coefficient taking into account the local hindrance effects (a) and a constant drag coefficient respectively. The dashed lines represent the transition between the different regimes, with a shaded area between the expected transition from turbulent to free-fall regime $r_c = 1$ and the one estimated by  \citet{CourrechDuPont2003} $r_c = 6$.}
\end{figure}

\begin{figure}
\centering
\includegraphics[width=\textwidth]{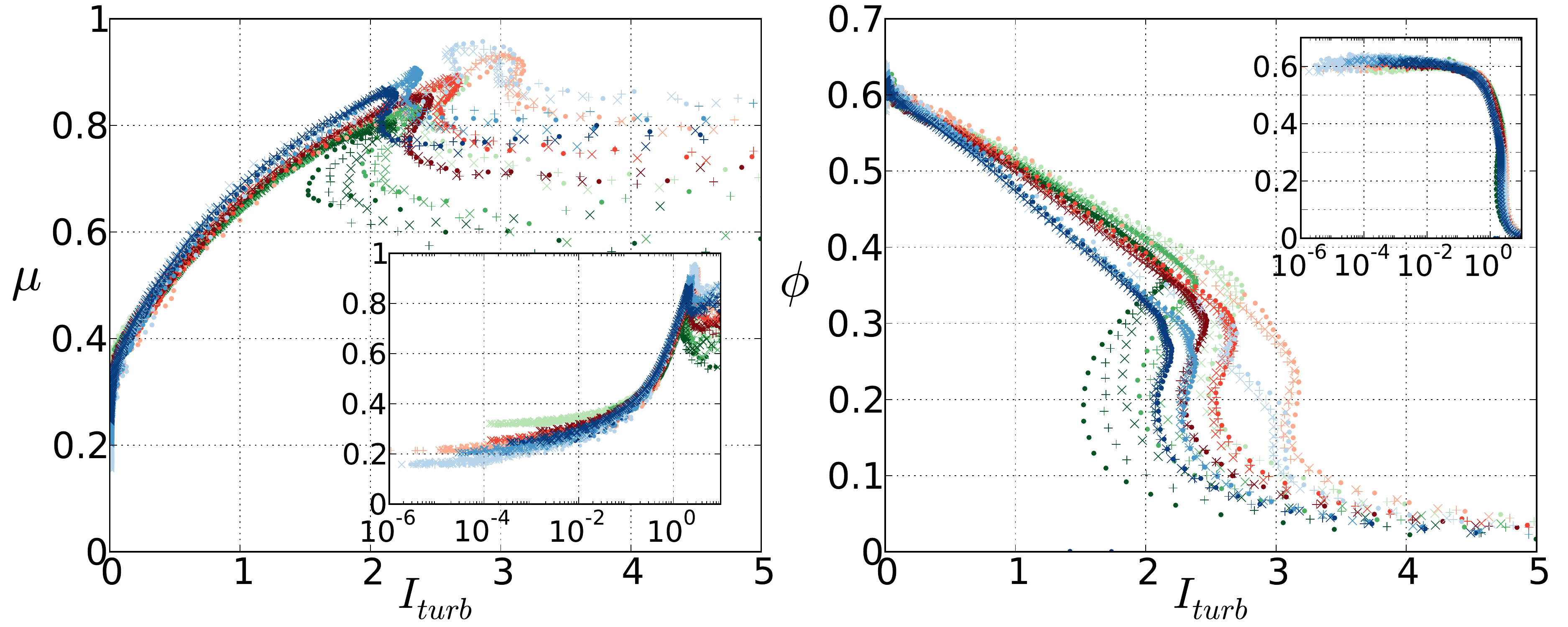} 
\caption{\label{figMuIturb} Shear to normal stress ratio $\mu = \tau^p/ P^p$ and solid volume fraction $\phi$ as a function of the turbulent inertial number $I_{turb} = \dot{\gamma} d\sqrt{\rho^f C_D/P^p}$, for all the cases presented in table \ref{tableRheoParam} with variation of Shields number, specific density and particle diameter.}
\end{figure}

\begin{figure}
  \centerline{\includegraphics[width=\textwidth]{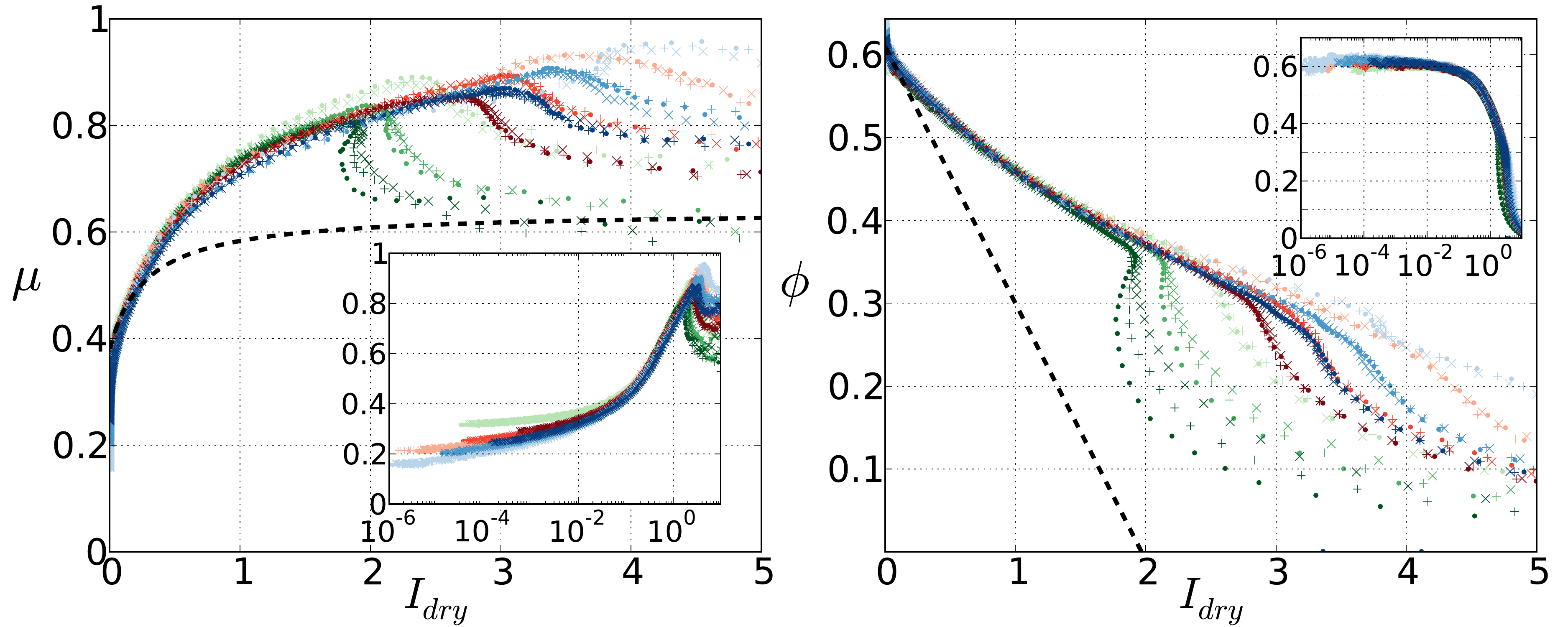}}
\caption{\label{muIAllLin} Shear to normal stress ratio $\mu = \tau^p/ P^p$ and solid volume fraction $\phi$ as a function of the dry inertial number $I_{dry} = \dot{\gamma}d \sqrt{\rho^p/P^p}$ for all the cases presented in table \ref{tableRheoParam} with variation of Shields number, specific density and particle diameter. The parameters of the simulation sampled and the corresponding symbols are shown in table \ref{tableRheoParam}. The dashed lines (--) represents the classical expression of $\mu(I)/\phi(I)$ (eq. \ref{muIeq} and \ref{phiIeq}) with the parametrization of \citet{DaCruz2005} and \citet{Jop2006}: $\mu_s = 0.38$, $\mu_2 = 0.64$, $I_0 = 0.279$, $a=0.31$.}
\end{figure}

Figure \ref{rSt}a shows the parameter variation in the $St$/$r$ plane.
According to the estimation of \citet{CourrechDuPont2003} ($r_c \sim 6$), all of the data belong to the turbulent regime and should show a collapse as a function of the turbulent inertial number. As can be seen from figure \ref{rSt}b, this would also be the case if the $r$ number was evaluated without taking into account the local hindrance effect. \\ 
Figure \ref{figMuIturb} and \ref{muIAllLin} show the shear to normal stress ratio and the solid volume fraction as a function of the turbulent inertial number and the dry inertial number respectively. Both figures exhibit a data collapse over the range $I \in [10^{-2},0.1]$. At higher inertial numbers however, the data split apart as a function of the turbulent inertial number, while the collapse persists up to $I_{dry} \sim 2$ for the dry inertial number case. Therefore, the configurations sampled are better described by the dry inertial number and belong at first order to the free-fall regime, in contradiction with the predicted transition from the free-fall to the turbulent regimes of \citet{CourrechDuPont2003}. One still notes a slight dependence on the specific density in figure \ref{muIAllLin}, suggesting a second-order influence of the fluid inertial rearrangement mechanism. This absence of clear transition from the free-fall to the turbulent regimes is consistent with recent global analysis of immersed granular collapse using a DNS-DEM model \citep{Izard2014}. Similarly to the transition from the viscous to the free-fall regime \citep{Trulsson2012}, one might expect the transition region to be described by a combination of the turbulent and dry inertial numbers. In the present paper, the second-order effects are neglected for simplicity and the dry inertial number will be adopted in the following. \\

Coming back to figure \ref{muIAllLin} and considering the semi-logarithmic scale insets, the solid volume fraction curves are seen to collapse down to the lowest inertial number sampled ($I_{dry}\sim 10^{-5}$). For the shear to normal stress ratio, the different curves split apart below  $I_{dry} \sim 10^{-2}$, following different branches depending on both the specific density and the Shields number of the run considered. No variation with the particle diameter is observed. In all cases, the stress ratio values in this region are below the expected static effective friction coefficient of the granular medium ($\mu_s \sim 0.38$ for monodisperse glass beads \citep{Andreotti2013}). This low-inertial-number region corresponds to the lower parts of the different simulations where a creeping regime is observed with the solid velocity following an exponential decrease \citep{Maurin2015,Houssais2015} similar to the one observed in dry granular flows on a heap \citep{Komatsu2001,Richard2008}. These two features are characteristic of non-local effects (e.g. \citet{Kamrin2012,Bouzid2013}), where the granular flow is influenced by the far field, i.e. by the top granular flow in the present configuration. Despite the interest for granular media and out-of-equilibrium configurations, the quasi-static part of bedload transport does not contribute significantly to the sediment transport rate and will not be investigated further in this paper. \\

Toward the upper limit, the data are seen to collapse up to $I_{dry}\sim 2$ for both the solid volume fraction and the shear to normal stress ratio (figure \ref{muIAllLin}). At higher inertial numbers, the different curves - corresponding to the different simulations - progressively exhibit a transition to a different behaviour characterised by a decrease of the shear to normal stress ratio, and a slope break on the solid volume fraction versus inertial number curves. \\

\begin{figure}
  \centerline{\includegraphics[width=\textwidth]{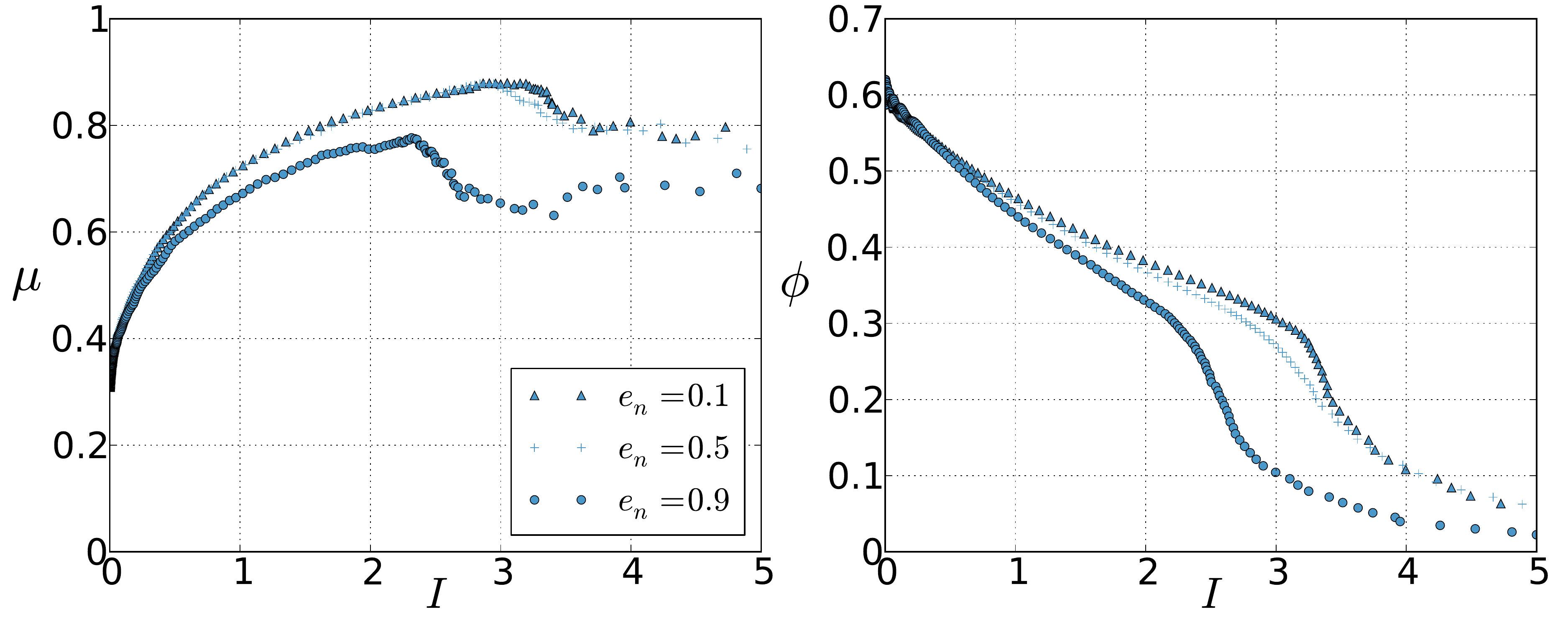}}
\caption{\label{figRestit} Effect of the restitution coefficient on the shear to normal stress ratio $\mu = \tau^p/ P^p$ and solid volume fraction $\phi$ as a function of the dry inertial number $I_{dry} = \dot{\gamma}d \sqrt{\rho^p/P^p}$ for a representative case (case r2d6s2 in table \ref{tableRheoParam}).}
\end{figure}

The observed decrease in the shear to normal stress ratio as a function of the inertial number is characteristic of the transition from dense to dilute granular behaviour \citep{Forterre2008}. In the present results, the position of the transition in terms of inertial number is particularly high ($I\sim 2-3$) and depends on the configurations sampled (figure \ref{muIAllLin}). At such a high inertial number, the restitution coefficient is seen to have a non-negligible effect on the rheological curves (see figure \ref{figRestit}). Focusing on the classical dry granular flow literature (e.g. \citet{DaCruz2005,Jop2006,Forterre2008,Jop2015}), the $\mu(I)$ approach is considered to break down above $I_{dry}\sim 0.5$ for glass beads, when collisional mechanisms comes into play \citep{DaCruz2004,Lois2006,Forterre2008}. Therefore, it is unable to explain both the observed variation of the transition position at constant restitution coefficient and the collapse observed in regions where collisional mechanisms contribute non-negligibly to the rheology. \\
Recently, similar persistence of dense granular flow behaviour at high inertial numbers has been observed in inclined plane configurations. \citet{Holyoake2012} observed experimentally dense granular flows up to inertial numbers $I_{dry} \sim 2$ in steady non-uniform dry granular flows of sand in a steep channel with lateral walls and a bumpy base. Moreover, analyzing granular roll wave instabilities using the DEM,  \citet{Borzsonyi2009} computed the local granular rheology and obtained a collapse of their rheological data up to $I_{dry} \sim 1$. In these two configurations as well as in the present study, the complexity of the granular flow considered can lead to differences with respect to the simple shear picture. In particular, the local analysis of spatially non-homogeneous granular flows made in the present paper and in \citet{Borzsonyi2009}, might lead to secondary gradient effects that are not taken into account in the $\mu(I)$ rheology. Therefore, the local character of the analysis and/or the specificity of the configuration seem to affect the persistence of a dense granular flow at high inertial numbers.\\
 As a consequence of the high-inertial-numbers collapse, the classical $\mu(I)/\phi(I)$ expressions (eq. \ref{muIeq} and \ref{phiIeq}) and parametrisation do not fit the present results well at high inertial numbers (see the fit of \citet{Jop2006} in figure \ref{muIAllLin}). Indeed, the collisional contribution becomes important in this region and one might need to extend the $\mu(I)/\phi(I)$ relationships as done by \citet{Holyoake2012} for the inclined plane configuration.\\

The results have evidenced the ability of the $\mu(I)$ framework to describe the granular flow rheology in bedload transport in the dense granular flow region. They have shown that the interstitial fluid does not influence the dense granular rheology importantly and that the fluid flow acts mainly as an external forcing in turbulent bedload transport. In addition, the analysis has underlined the variety and the complexity of the granular behaviours observed in bedload transport, which challenge the existing formulation of the $\mu(I)$ rheology. \\

\section{Granular rheology in bedload transport}
\label{graRheoBed}

The rest of this paper focuses on the analysis of turbulent bedload transport, considering realistic conditions typical of gravel-bed rivers. Therefore, the specific density is taken as $\rho^p/\rho^f -1 = 1.5$ and additional DEM simulations are performed at Shields numbers down to the onset of motion. The analysis of the DEM results allows us to propose a $\mu(I)$-based granular rheology for turbulent bedload transport. The relevance of the proposed granular rheology is further tested in a second time, using the Eulerian-Eulerian model presented in section \ref{modelFormulation}. 

\subsection{Parametrisation of the $\mu(I)$ rheology} 
\label{paraMuI}

\begin{figure}
\centering
  \includegraphics[width=\textwidth]{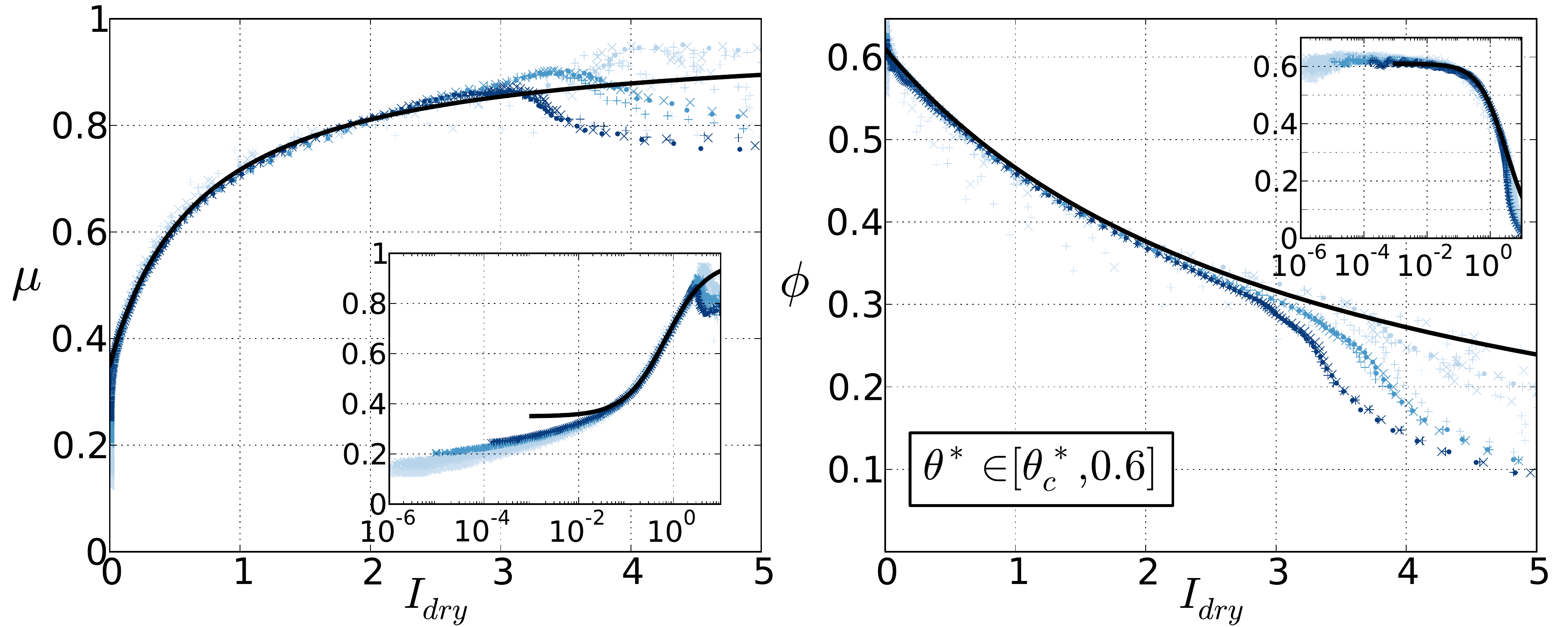} 
\caption{\label{figFitmuI} Shear to normal stress ratio and solid volume fraction as a function of the dry inertial number, for realistic bedload transport simulations with specific density $\rho^p/\rho^f -1 = 1.5$, Shields number from incipient motion to $0.7$, and particle diameter of $d = 3$mm,  $d = 6$mm, and  $d = 12$mm. The black lines represent the best fit obtained with equations \ref{muIeq} and \ref{phiIModif} ($\mu_1 = 0.38$, $\mu_2 = 0.64$, $I_0 = 0.279$, $\phi^{max} = 0.61$, and $b = 0.31$). The symbols colors are function of the simulation Shields number: the darkest the color, the highest the Shields number.}
\end{figure}

Figure \ref{figFitmuI} shows the granular shear to normal stress ratio and the solid volume fraction as a function of the inertial number, for simulations with Shields numbers from $\theta^* = 0.04$ to $\theta^* = 0.6$ and for particle diameters $d \in \{ 3 ; 6 ; 12\}$ mm. In these new configurations closer to realistic bedload transport, most of the data are seen to collapse for both the solid volume fraction and the granular stress ratio up to high inertial numbers. Above $I \sim 3$ and under $I\sim 10^{-2}$, the curves split apart similarly to the general case. No dependence upon the particle diameter is observed over the whole range of inertial numbers, pointing out the negligible effect of the $Re_p$ and $St$ numbers in the problem. The lowest-Shields-number cases (lightest points) exhibit some specific behaviour with oscillations, lower shear to normal stress ratio and solid volume fraction at given inertial numbers. These cases correspond to the sharpest transitions between a quasi-static granular bed and a few particles in motion on top of it. Therefore, in these configurations the continuous assumption is not strictly valid and the limits of the rheological description are reached. \\

Overriding these limitations close to the critical Shields number and assuming that the high-inertial-numbers granular behaviour is not dominant in bedload transport, the $\mu(I)$ constitutive laws are pragmatically best-fitted on the Eulerian-Lagrangian simulation results over the range of inertial numbers $I_{dry} \in [10^{-2},3]$. The classical solid volume fraction expression (eq. \ref{phiIeq}) predicts negative values at high inertial numbers and is unable to describe the present results. Consequently, the expression is modified and is chosen following \citet{Aussillous2013} and \citet{RevilBaudard2013} as
\begin{equation}
\phi(I) = \frac{\phi^{max}}{1 + b I}.
\label{phiIModif}
\end{equation}
The best fits of equations \ref{muIeq} and \ref{phiIModif} on the data give the following parameters: $b = 0.31$, $\mu_s = 0.35$, $\mu_2 = 0.97$ and $I_0 = 0.69$, with $\phi^{max}$ imposed to take the value measured in the Eulerian-Lagrangian simulations: $\phi^{max}= 0.61$. Developing the solid volume fraction expression around $I=0$, the value of $b$ obtained is consistent with both the parametrisation of \citet{DaCruz2005} in the dry regime and the one from \citet{Boyer2011} and \citet{Trulsson2012} in the viscous regime. The values of $\mu_2$ and $I_0$ are logically higher than in the original $\mu(I)$ rheology, as no real saturation of the stress ratio is observed in our data. The best fits describe very well the results obtained within the collapsing range of $I_{dry}$, as can be seen in figure \ref{figFitmuI}. Outside this interval, they do not reproduce the non-local effects at low inertial numbers ($I_{dry}<10^{-2}$) and the drop in both the shear to normal stress ratio and the solid volume fraction for $I_{dry}>3$. \\
Assuming the dense granular behaviour to be dominant, the two relationships (eq. \ref{muIeq}-\ref{phiIModif}) together with the values determined from the best fits are proposed as a $\mu(I)$-based granular rheology for turbulent bedload transport.

\subsection{Eulerian-Eulerian simulations}

In order to evaluate the relevance of the proposed granular rheology it has been implemented in the Eulerian-Eulerian model presented in section \ref{modelFormulation}. Simulations have been performed using exactly the same parameters for both the Eulerian-Lagrangian and the Eulerian-Eulerian models ($\rho_p/\rho_f-1=1.5$ ; $d=6$ mm, $\alpha=0.05$, $h$). The particle diameter has been varied ($d\in{3,6,12}$mm) but does not influence the results so that only the case $d = 6$mm is presented for clarity. \\

\begin{figure}
\centering
  \includegraphics[width=\textwidth]{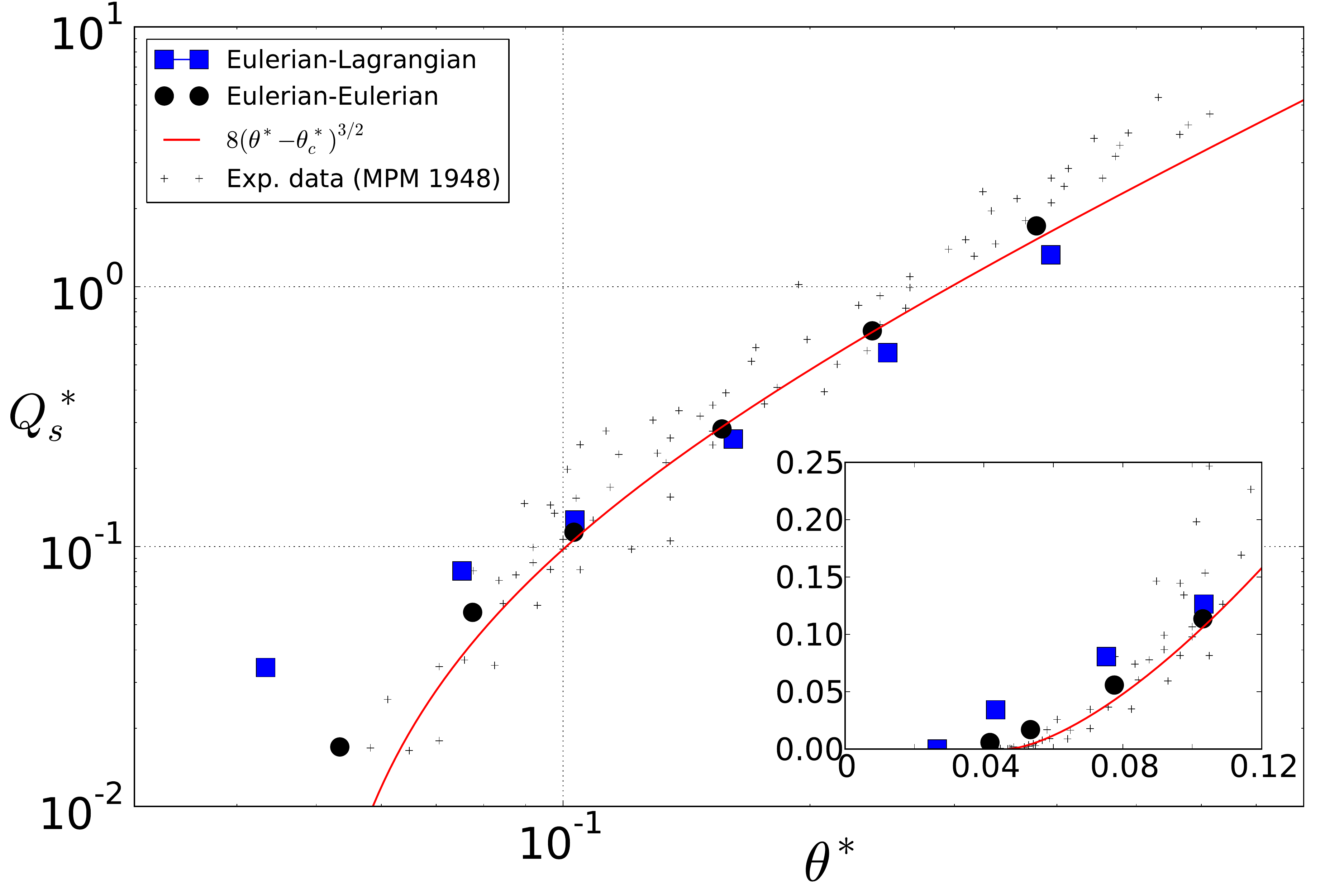} 
\caption{\label{compQsTheta} Dimensionless sediment transport rate as a function of the Shields number for two-phase continuous simulations (Eulerian-Eulerian) and coupled fluid-discrete element simulations (Eulerian-Lagrangian), with realistic parameters for turbulent bedload transport: $\rho^p/\rho^f -1 = 1.5$ and $d =6$mm. The black crosses and the red line represent the data and the fit of \citet{MPM1948} respectively.}
\end{figure}
\begin{figure}
\centering
  \includegraphics[width=\textwidth]{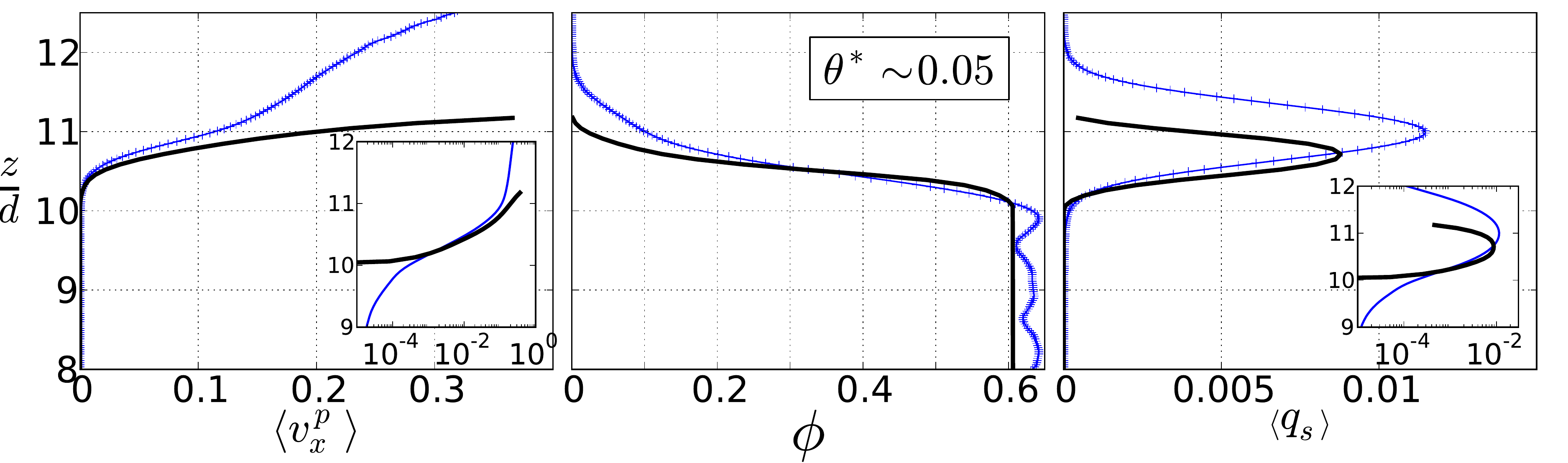} 
  \includegraphics[width=\textwidth]{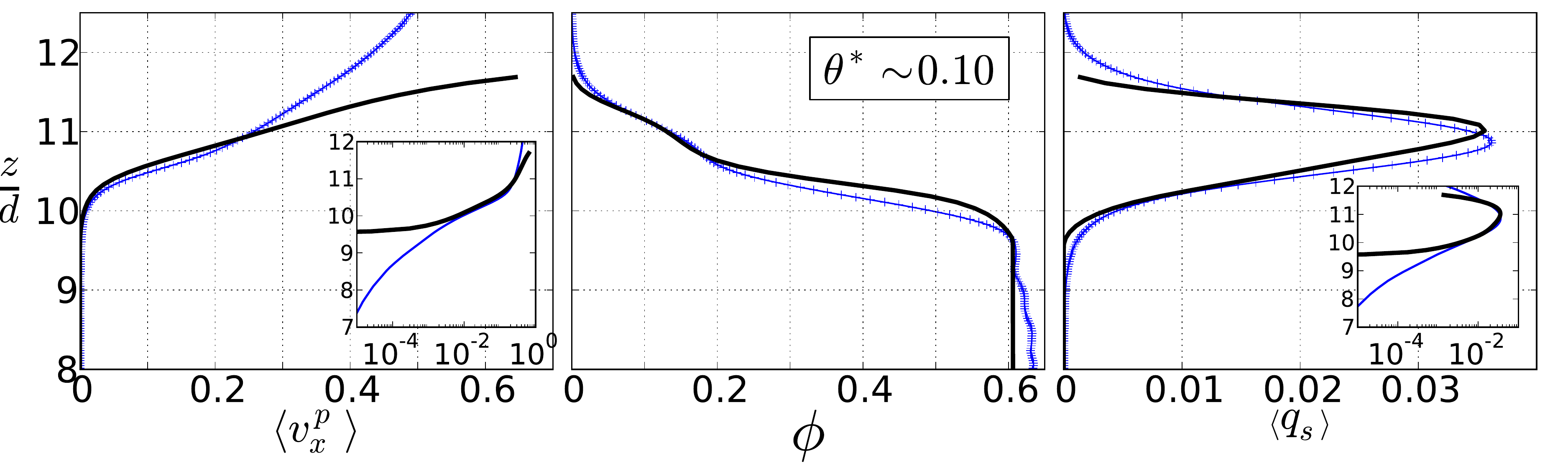} 
  \includegraphics[width=\textwidth]{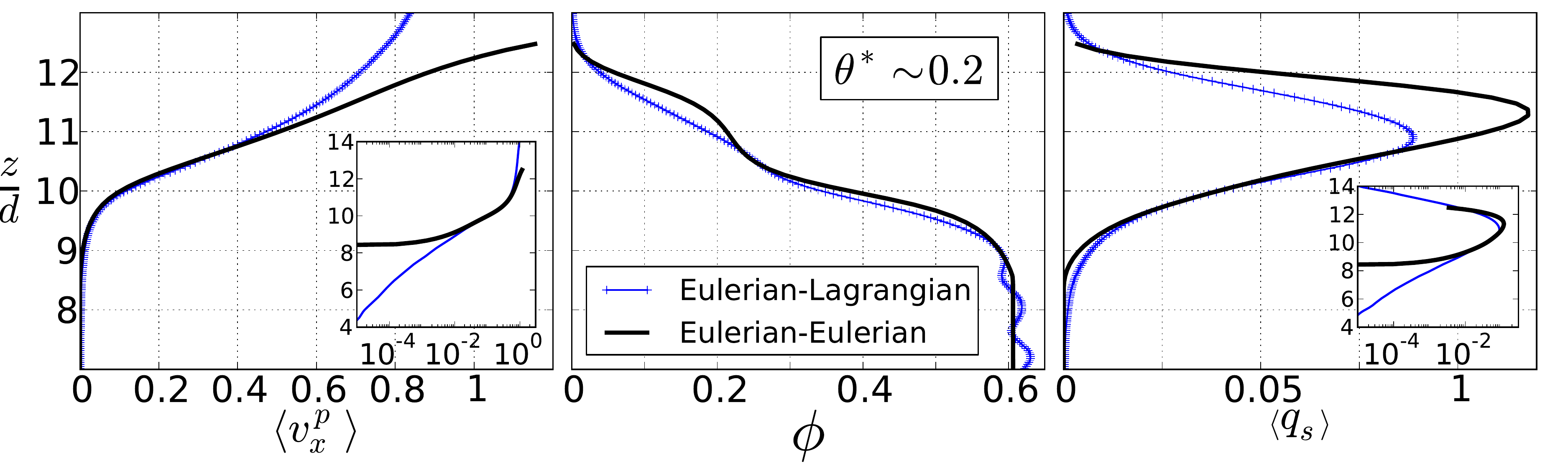} 
\caption{\label{compSolidPro} Solid depth profile comparison between the two-phase continuous model (Euler-Euler) and the coupled fluid-discrete element model (Euler-Lagrange) for three different Shields number $\theta^*$ with $d= 6mm$ and $\rho^p/\rho^f -1= 1.5$. Solid velocity $<v^p_x>^s$ and transport rate density $<q_s>^s$ are expressed in $m/s$ while the solid volume fraction is dimensionless. The fluid mechanics convention is used.}
\end{figure}

Figure \ref{compQsTheta} presents the dimensionless sediment transport rate versus the Shields number for the Eulerian-Eulerian model, the Eulerian-Lagrangian model and the experimental data and empirical law of \citet{MPM1948}. A good agreement with the experimental data is obtained using both models. The proposed granular rheology enables the Eulerian-Eulerian model to accurately reproduce the critical Shields number ($\theta^*_c \sim 0.04$) and the global trend over a wide range of Shields numbers ($\theta^* \in [\theta^*_c,0.7]$). Quite surprinsingly, the Eulerian-Eulerian model better predicts the smooth transition around the critical Shields number, in a region where the continuous assumption breaks down. Conversely, at high Shields numbers the increased importance of the dilute region, which is not well described by the proposed rheology (section \ref{paraMuI}), induces a slight discrepancy between the Eulerian-Eulerian and Eulerian-Lagragian models.\\

To go further in the analysis, the comparison is extended to the granular depth profiles by decomposing the sediment transport rate into the integral of the sediment transport rate density $\left<q_s\right>(z)=\left<v_x\right>^s(z) \ \phi(z)$ \citep{Bagnold1956}:
\begin{equation}
Q_s = \int_0^\infty{\left<q_s\right>(z) dz} =  \int_0^\infty{\left<v_x\right>^s(z) \ \phi(z) dz}.
\end{equation}
Figure \ref{compSolidPro} shows the comparison of the solid volume fraction, the solid velocity and the transport rate density profiles obtained with the Eulerian-Eulerian and Eulerian-Lagrangian models for three different Shields numbers $\theta^* =  \{ 0.05 ; 0.1 ; 0.2\}$. The Eulerian-Eulerian model reproduces accurately the granular depth profiles obtained with the Eulerian-Lagrangian model for solid volume fractions approximately higher than 0.3 ($\phi \gtrsim 0.3$). This value roughly corresponds to the domain for which the proposed granular rheology describes the Eulerian-Lagrangian results very well (see figure \ref{figFitmuI}). Only the lower quasi-static velocity profile - negligible in terms of sediment transport rate - is not reproduced, as non-local effects have not been taken into account in the granular rheology. In the dilute region ($\phi \lesssim 0.3$), the particle velocity is overestimated by the Eulerian-Eulerian model for the three different Shields numbers, showing an increasing discrepancy toward the lower solid volume fraction region. This is associated with an under-estimation of the solid volume fraction close to incipient motion and an over-estimation at high Shield numbers. These behaviours point out both the ill-posedness of the $\mu(I)$ rheology as the solid volume fraction goes to zero and the inability of the proposed rheology to describe the dilute region accurately. Despite these limitations, the good lower boundary condition given by the accurate description of the dense granular flow constrains the dilute behaviour and enables a good description of the sediment transport rate over a wide range of Shields numbers with the Eulerian-Eulerian model. \\

The present results show that a $\mu(I)$ rheology is able to reproduce the main features of turbulent bedload transport in terms of sediment transport rate and dense granular depth profiles, in steady uniform configurations. Considering the much lower computational cost of Eulerian-Eulerian simulations compared with Eulerian-Lagrangian ones, it will allow an important step to be made in the upscaling process towards applications. In this perspective, the fluid phase description might have to be upgraded, using a two-equation turbulence model as $k-\epsilon$ or $k-\omega$ \citep{Hsu2004,Amoudry2014,Lee2016}, or eddy resolving simulations.

\section{Summary and conclusion}
The granular rheology in idealised turbulent bedload transport has been studied using a coupled fluid-discrete-element model. Computing the granular stress tensor locally as a function of the depth and performing simulations for various Shields numbers, specific densities and particle diameters, the data have been shown to collapse over an important range of inertial numbers, demonstrating the relevance of the $\mu(I)$ rheology for turbulent bedload transport description. No clear transition from the free-fall to the turbulent regime of the $\mu(I)$ rheology has been observed, supporting recent numerical observations and suggesting that the interstitial fluid does not importantly influence the granular rheology in turbulent bedload transport. In addition, consistently with recent literature, the data have been observed to collapse in terms of both the solid volume fraction and the shear to normal stress ratio, up to inertial numbers as high as $I \sim 2$. The persistence of a dense granular flow behaviour at such a high inertial number together with the observed progressive transition from dense to dilute behaviour challenges the classical conceptions and parametrizations of the $\mu(I)$ rheology, and opens interesting perspectives for a better understanding of granular media rheology at high inertial numbers. Beyond these observations, a $\mu(I)$-based granular rheology has been fitted to the present Eulerian-Lagrangian bedload transport simulations. The proposed rheology has been sucessfully tested using a Eulerian-Eulerian model. It has been shown to accurately reproduce the dense granular depth profiles and the classical experimental results in terms of sediment transport rate from the onset of motion up to a Shields number of $\theta^* \sim 0.7$. \\
Further work is needed to better understand the high-inertial-numbers behaviour and extend the granular rheology to the description of the dilute region. At this stage it is not clear whether the $\mu(I)$ rheology could be extended to handle this regime transition. Besides, the proposed rheology can already be used in three dimensional Eulerian-Eulerian numerical models and represents a step in the upscaling process from particle-scale simulations toward full-scale problem modelling. 

\section*{Aknowledgement}
We are greatful to the three anonymous reviewers for relevant and constructive questions, and to Ashley Dudill for English corrections. RM would like to thank Thierry Faug for fruitful discussions.
This research was supported by Irstea (formerly Cemagref), the labex OSUG@2020 and the French Institut National des Sciences de l'Univers programs EC2CO-BIOHEFECT and EC2CO-LEFE MODSED.

\appendix

\section{Momentum balance derivation}
\label{appendixMomBal}
In order to validate the stress tensor profile computed, the momentum balance is analyzed in the framework of the continuous two-phase equations \citep{Jackson2000}. In the present appendix, the equations are derived and integrated to show the exact terms computed in the body of the paper.  Using the steady and uniform character of the problem considered, the momentum balances for both the fluid and the granular phases read along the streamwise direction \citep{Jackson2000}
\begin{equation}
0 = \frac{\partial S_{xz}^f}{\partial z} + \frac{\partial R_{xz}^f}{\partial z} + \rho^f (1-\phi) g \sin \alpha - n \left<{f_f^p}_x\right>^p,
\label{fluidXZch5}
\end{equation}
\begin{equation}
0 = \frac{\partial \tau_{xz}^p}{\partial z} + \rho^p \phi  g \sin \alpha + n \left<{f_f^p}_x\right>^p.
\label{partXZch5}
\end{equation}
Combining these two equations together, the mixture momentum balance can be written as
\begin{equation}
0 = \frac{\partial S_{xz}^f}{\partial z} + \frac{\partial R_{xz}^f}{\partial z} + \frac{\partial \tau_{xz}^p}{\partial z} + (\rho^p \phi + (1-\phi)\rho^f) g \sin \alpha.
\label{mixtureXZch5}
\end{equation}
In order to study the stress repartition, the equation is integrated between a given position $z$ in $h>z>0$  and the water free-surface elevation $h$, where the viscous, turbulent and particle shear stresses vanish: $S_{xz}^f(h) = R_{xz}^f(h) = \tau_{xz}^p(h) = 0$. It leads to the following formulation: 
\begin{equation}
0 = -S_{xz}^f(z) - R_{xz}^f(z) - \tau_{xz}^p(z) + g \sin \alpha\left[\rho^f (h-z) + (\rho^p-\rho^f) \int_z^h{\phi dz}\right] .
\label{mixtureXZ_INT}
\end{equation}
From one simulation, it is possible to evaluate all of the terms of the equation at each given elevation. Provided that the stress tensor formulation is appropriate and the system is at equilibrium, the equality should be satisfied.

\end{document}